\documentstyle[12pt]{article}
\title{\bf Casimir effect for concentric spheres in de Sitter spacetime with signature change}
\author{F. Darabi \thanks{E-mail:f.arabi@azaruniv.edu}
\\ {\small Department of Physics, Azarbaijan University of Tarbiat Moallem,
Tabriz, 53714-161 Iran.}\\
{\small Research Institute for Astronomy and Astrophysics of Maragha
- Maragha, 55134-441  Iran. } }
\begin{document}
\maketitle \vspace{15mm}
\begin{abstract}
We revisit the Casimir effect for two concentric spherical shells in
de Sitter background with a new geometric configuration, namely
Euclidean signature between and Lorentzian signature outside the
spheres with different cosmological constants, for a massless scalar
field satisfying Dirichlet boundary conditions on the spheres. It is
shown that an extra constant pressure emerges due to this signature
changing configuration. Some interesting aspects of this extra term
are then discussed.
\end{abstract}
% \begin{document}
\newpage
% \vspace*{10mm}
\section{Introduction}

The Casimir effect is regarded as one of the most striking
manifestation of vacuum fluctuations in quantum field theory. The
presence of reflecting boundaries alters the zero-point modes of a
quantized field, and results in the shifts in the vacuum expectation
values of quantities quadratic in the field, such as the energy
density and stresses. Therefore, the Casimir effect can be viewed as
the polarization of the vacuum by boundary conditions or geometry.
In particular, vacuum forces arise acting on the constraining
boundaries. The particular features of these forces depend on the
nature of the quantum field, the type of spacetime manifold and its
dimensionality, the boundary geometries and the specific boundary
conditions imposed on the field. Since the original work by Casimir
in 1948 \cite{Casi48} many theoretical and experimental works have
been done on this problem
\cite{db,Most97,Plun86,Lamo99,Bord01,Kirs01,Schw,setare}.

The time dependence of boundary conditions or geometries, the
so-called dynamical Casimir effect, is also a new element which has
to be taken into account. In particular, in \cite{set} the Casimir
effect has been calculated for a massless scalar field satisfying
Dirichlet boundary conditions on the spherical shell in de Sitter
space. The Casimir stress is calculated for inside and outside of
the shell with different backgrounds corresponding to different
cosmological constants.

On the other hand, signature changing spacetimes have recently been
of particular importance as specific geometries with interesting
physical effects. The initial idea of signature change is due to
Hartle, Hawking and Sakharov \cite{HHS} which makes it possible to
have a spacetime with Euclidean and Lorentzian regions in quantum
gravity. It has been shown that the signature change may happen even
in classical general relativity \cite{CSC}. The issue of propagation
of quantum fields on signature-changing spacetimes has also been of
some interest \cite{D}. For example, Dray {\it et al} have shown
that the phenomenon of particle production can happen for
propagation of scalar particles in spacetime with heterotic
signature. They have also obtained a rule for propagation of
massless scalar fields on a two dimensional spacetime with signature
change.

The Casimir effect has also recently been studied in a signature
changing two dimensional spacetime, having a cylindrical topology
with a narrow Euclidean region encircled by outside Lorentzian
signature, and shown that there is a non-vanishing pressure on the
hypersurface of signature change \cite{FM}. The Casimir stress on a
spherical shell in de Sitter signature changing background for
massless scalar field satisfying Dirichlet boundary conditions on
the shell is also calculated \cite{MF}. Motivated by this new
element in studying the Casimir effect we have paid attention to
study such a non-trivial effect in a model of two cocentric
spherical shells in de Sitter space with different cosmological
constants and metric signatures, namely Euclidean signature between
and Lorentzian one outside the spheres. The Casimir stress on two
cocentric spherical shells in de Sitter spacetime with fixed
Lorentzian background has already been obtained in \cite{Coc}. In
this paper, we will generalize this result to a signature changing
background.

  \section{Casimir effect for concentric spheres in flat space-time}

  In this section we shall use the results of \cite{Sahar1}.
  Consider two concentric spherical shells with zero thickness and
   radii $a$ and $b$, $a<b$. Consider now the Casimir force due to
  fluctuation of a free massless scalar field satisfying Dirichlet
  boundary conditions on the spherical shells in Lorentzian
  space-time.
  The vacuum force per unit area of the inner sphere is given by
  \begin{equation}\label{Feq1}
  F_{a}(a,b)=F(a)-P_{a}(a,b,a),
  \end{equation}
  where $F(a)$ is the force per unit area of a single sphere with
  radius $a$, and $P_{a}(a,b,a)$ is due to the existence of the
  second sphere (interaction force). Similarly, the vacuum force
  acting per unit area of outer sphere is
  \begin{equation}\label{Feq2}
  F_{b}(a,b)=F(b)+P_{b}(a,b,b).
  \end{equation}
  Therefore, the vacuum force per unit area of a single sphere is the sum of Casimir
  forces $F_{in}$ and $F_{out}$ for inside and outside of the
  shell.
  \begin{equation}
  F(a)=F_{in}(a)+F_{out}(a),\hspace{1cm}
  F(b)=F_{in}(b)+F_{out}(b).
  \end{equation}
  Casimir forces inside and outside of the shell are divergent individually \cite{set2},
  but when we add interior and exterior forces to each other,
  divergent parts cancel each other out. Interaction forces $P_{a}(a,b,a)$ and
  $P_{b}(a,b,b)$ are finite and for Dirichlet boundary condition, are given by
  \begin{equation}\label{peq1}
  P_{a}(a,b,a)=\frac{-1}{8\pi^{2}a^{3}}\sum_{l=0}^{\infty}(2l+1)\int_{0}^{\infty}
  dz\frac{K_{\nu}^{(b)}(bz)/K_{\nu}^{(a)}(az)}
  {K_{\nu}^{(a)}(az)I_{\nu}^{(b)}(bz)-K_{\nu}^{(b)}(bz)I_{\nu}^{(a)}(az)},
  \end{equation}
 \begin{equation}\label{peq2}
  P_{b}(a,b,b)=\frac{-1}{8\pi^{2}b^{3}}\sum_{l=0}^{\infty}(2l+1)\int_{0}^{\infty}
  dz\frac{I_{\nu}^{(a)}(az)/I_{\nu}^{(b)}(bz)}
  {K_{\nu}^{(a)}(az)I_{\nu}^{(b)}(bz)-K_{\nu}^{(b)}(bz)I_{\nu}^{(a)}(az)},
  \end{equation}
  where $I_{\nu}$ and $K_{\nu}$ are modified Bessel function.

\section{Casimir effect for a sphere in de Sitter space}

We will consider a conformally coupled massless scalar field $\phi$
satisfying the following Dirichlet boundary condition on a spherical
shell in de Sitter space
\begin{equation}
\phi(x=a)=0 , \label{boun}
\end{equation}
where $a$ is the radius of spherical shell. The corresponding field
equation has the form
\begin{equation}
(\nabla_{\mu}\nabla^{\mu}+\xi R)\phi=0, \hspace{1cm}
\nabla_{\mu}\nabla^{\mu}=\frac{1}{\sqrt{-g}}\partial_{\mu}(\sqrt{-g}g^{\mu\nu}\partial_{\nu}),
\label{eqmo}
\end{equation}
where $\nabla_{\mu}$ is the covariant derivative operator, and $R$
is the Ricci scalar for de Sitter space. In the conformally coupled
case, the corresponding stress-energy tensor is defined as \cite{db}
\begin{equation}
T_{\mu\nu}=(1-2\xi)\partial_{\mu}\phi\partial_{\nu}\phi+(2\xi-1/2)g_{\mu\nu}\partial^{\lambda}\phi\partial_{\lambda}
\phi-2\xi\phi\nabla_{\mu}\nabla_{\nu}\phi+\frac{1}{12}g_{\mu
\nu}\phi\nabla_{\lambda}\nabla^{\lambda}\phi, \label{enmom}
\end{equation}
where $\xi=\frac{1}{6}$. To make maximum use of the flat-space
calculations we use de Sitter metric in conformally flat form
\begin{equation}
ds^{2}=\Omega^{2}(\eta)[d\eta^{2}-\sum_{i=1}^{3}(dx^{i})^{2}],
\label{6}
\end{equation}
where $\Omega(\eta)=\frac{\alpha}{\eta}$ and $\eta$ is the conformal
time \cite{db}
\begin{equation}
-\infty <  \eta < 0, \label{7}
\end{equation}
which covers half of de Sitter space conformal to a portion of
Minkowski space. Notice that we assume the same coordinate system
and the same range for the values of coordinates in two conformally
related metrics. In particular, if we take from the one side a
spherical shell in the Minkowski spacetime with the time variable
$-\infty < \eta < \infty$, in the corresponding problem in de Sitter
spacetime the range for $\eta$ should be the same. However, since
the metric (\ref{6}) is symmetric under $\eta \rightarrow -\eta$ and
we are interested in a cosmology with forward in time evolution, we
then consider the range $0 < \eta < \infty$ for this metric to cover
the other half.

  The quantization of a scalar field on background of the metric
  (\ref{6}) is straightforward. Let $\{\phi_{k}(x), \phi^{\star}_{k}(x)\}$
  be a complete set of orthonormalized positive and negative
  frequency solutions to the field equation (\ref{eqmo}), obeying
  boundary condition (\ref{boun}). By expanding the field operator
  over these eigenfunctions, using the standard commutation rules
  and the definition of the vacuum state for the vacuum
  expectation values of the energy-momentum tensor one obtains
  \cite{db}
\begin{equation}
\langle0|T_{\mu}^{\nu}|0\rangle=\sum_{k}T_{\mu}^{\nu}\{\phi_{k},
\phi_{k}^{*}\}, \label{19}
\end{equation}
where $|0>$ is the amplitude for the corresponding vacuum state, and
the bilinear form $T_{\mu}^{\nu}\{\phi,\phi^{*} \}$ on the right is
determined by the classical energy-momentum tensor (\ref{enmom}). In
the problem under consideration we have a conformally trivial
situation, namely a conformally invariant field on background of the
conformally flat spacetime. Instead of evaluating Eq.(\ref{19})
directly on background of the curved metric, the vacuum expectation
values can be obtained from the corresponding flat spacetime results
for a scalar field $\phi$ by using the conformal properties of the
problem under consideration. Under the conformal transformation
$\tilde g_{\mu\nu}=\Omega^{2}g_{\mu\nu}$ the $\phi$ field will be
changed by the rule
\begin{equation}\label{fieltra}
\tilde\phi(x)=\Omega^{-1}{\phi}(x).\label{*}
\end{equation}
Given two conformally related metrics $g_{\mu\nu}$ and $\tilde
g_{\mu\nu}$, the corresponding energy-momentum tensor for
conformally coupled situations are related as \cite{db}
\begin{equation}
\langle0|T^{\nu}_{\mu}[\tilde
g_{kl}]|0\rangle|_{ren}=(g/\tilde{g})^{1/2}\langle0|T^{\nu}_{\mu}[g_{kl}]|0\rangle|_{ren}-
\frac{1}{2880\pi^2}[{\frac{1}{6}}\vspace{2mm} ^{(1)}\tilde
H^{\nu}_{\mu}- ^{(3)}\tilde H^{\nu}_{\mu}], \label{15}
\end{equation}
where $g$ and $\tilde{g}$ are the determinants of the corresponding
metrics. Now $\langle0|T^{\nu}_{\mu}[g_{kl}]|0\rangle|_{ren}$, is
the regularized energy-momentum tensor for a conformally coupled
scalar field in the case of a spherical shell in flat spacetime. The
second term in (\ref{15}) is the vacuum polarization due to the
gravitational field, without any boundary condition. The functions
$^{(1,3)}\tilde H^{\nu}_{\mu}$ are some combinations of curvature
tensor components \cite{db}.
  The radial Casimir force per unit area $\frac{F}{A}$ on the
  sphere, called Casimir stress, is obtained from the radial-radial component of the vacuum
  expectation value of the stress-energy tensor:
  \begin{equation}
  \frac{F}{A}=\langle0|T_{(in)}{^{r}_{r}}-T_{(out)}{^{r}_{r}}|0\rangle|_{x=a}.
  \label{4}
  \end{equation}

  \section{Casimir effect for concentric spheres in de sitter space with different signatures }

Consider the system of two concentric spherical shells with zero
thickness and with radii $a$ and $b$, $(a < b)$ in de Sitter
space. We assume different vacua in between and outside the
spheres, corresponding to different $\alpha_{bet}$ and
$\alpha_{out}$. We also assume Euclidean signature for the region
between the spheres, and Lorentzian signature for the region
outside the spheres. We will use the de sitter metric in
conformally flat form (\ref{6}). The relation between parameter
$\alpha$ and cosmological constant $\Lambda$ is given by
  \begin{equation}\label{const}
  \alpha^{2}=\frac{3}{\Lambda}.
  \end{equation}
The energy for each single sphere is then obtained as
\begin{equation}\label{rEeq01}
\bar
E(a)=\frac{\eta^{2}}{6a}[(c_{1}\Lambda_{out}+c_{2}\Lambda_{bet})+
\frac{c'_{1}}{\varepsilon}(\Lambda_{out}-\Lambda_{bet})],
\end{equation}
\begin{equation}\label{rEeq02}
\bar E(b)=\frac{\eta^{2}}{6b}[(c_{1}\Lambda_{bet}+
c_{2}\Lambda_{out})+
\frac{c'_{1}}{\varepsilon}(\Lambda_{bet}-\Lambda_{out})],
\end{equation}
for which the renormalization leads to the Casimir energies
\cite{Coc}
\begin{equation}\label{rEeq1}
\bar
E_{ren}(a)=\frac{\eta^{2}}{6a}(c_{1}\Lambda_{out}+c_{2}\Lambda_{bet}),
\end{equation}
\begin{equation}\label{rEeq2}
\bar
E_{ren}(b)=\frac{\eta^{2}}{6b}(c_{1}\Lambda_{bet}+c_{2}\Lambda_{out}).
\end{equation}
where $c_1 = 0.008873, c_2 = -0.003234, c'_1 = 0.001010$. Now we
use the following relation for the stress on a shell in the
Lorentzian metric
\begin{equation}\label{eqEF}
\left(\frac{\bar F(a)}{A}\right)_L=\frac{-1}{4\pi
a^{2}}\frac{\partial \bar{E}}{\partial a}.
\end{equation}
In the signature changing case considered here we have
$\alpha_{in} \equiv \alpha_{out}$ and $\alpha_{out} \equiv
\alpha_{bet}$ for the sphere with radius $a$, and $\alpha_{in}
\equiv \alpha_{bet}$ and $\alpha_{out} \equiv \alpha_{out}$ for
the sphere with radius $b$. Correspondingly we have
\begin{equation}
\left(\frac{\bar F(a)}{A}\right)_{L-E}=
[\langle0|T_{r}^{r}|0\rangle|^{L}_{out}-\langle0|T_{r}^{r}|0\rangle|^{E}_{bet}].
\label{18}
\end{equation}
\begin{equation}
\left(\frac{\bar F(b)}{A}\right)_{E-L}=
[\langle0|T_{r}^{r}|0\rangle|^{E}_{bet}-\langle0|T_{r}^{r}|0\rangle|^{L}_{out}].
\label{18'}
\end{equation}
The scalar field $\phi(\vec{x}, \eta)$ in the Lorentzian de Sitter
space satisfies
\begin{equation}
(\Box+\xi R)\phi(\vec{x}, \eta)=0, \label{21}
\end{equation}
where $\Box$ is the Laplace-Beltrami operator for de Sitter metric,
and $\xi$ is the coupling constant. For conformally coupled field in
four dimension $\xi=\frac{1}{6}$, and R , the Ricci scalar
curvature, is given by
\begin{equation}
R=12\alpha^{-2}.
\end{equation}
Taking into account the separation of variables as
\begin{equation}
\phi_L(r,\theta,\eta)=A(r)B(\theta)T_L(\eta), \label{22}
\end{equation}
for the Lorentzian region with
\begin{equation}
T_L(\eta)=\frac{1}{\sqrt{2\omega(2\pi)^3}}\exp^{-i\omega\eta},
\label{23}
\end{equation}
the corresponding Euclidean $\eta$-dependent term takes on the
form
\begin{equation}
T_E(\eta)=\frac{1}{\sqrt{2\omega(2\pi)^3}}\exp^{-\omega\eta},
\label{24}
\end{equation}
so that the scalar field be normalized in $\eta$. It is easily
expected that the only difference between
$\left(\frac{\bar{F}}{A}\right)_{L-E}$,
$\left(\frac{\bar{F}}{A}\right)_{E-L}$ in one hand, and
$\left(\frac{\bar{F}}{A}\right)_{L}$ on the other hand, which was
obtained in \cite{set}, lies in the calculations for $T_L(\eta)$
and $T_E(\eta)$. Based on the following mode expansion
\begin{equation}
\phi=\sum_{i}(a_i^-\varphi_i^-+a_i^+\varphi_i^+),
\end{equation}
and normal ordering $\langle0|a_i^- a_i^+ |0\rangle=1$,  the
detailed calculations, using Eqs.(\ref{19}),(\ref{23}) and
(\ref{24}) in Eqs.(\ref{18}) and (\ref{18'}), lead to the stresses
on each single shell due to the boundary conditions
\begin{equation}\label{rFeq1}
\left(\frac {\bar F(a)}{A}\right)_{L-E}=\frac{-1}{4\pi
a^{2}}\frac{\partial\bar E(a)}{\partial a}=\frac{\eta^{2}}{24\pi
a^{4}}(c_{1}\Lambda_{out}+c_{2}\Lambda_{bet})\left(1+\frac{1}{64\pi^5\eta^2}(\Lambda_{out}^{-1}-
\Lambda_{bet}^{-1})\zeta(2)\right),
\end{equation}
\begin{equation}\label{rFeq2}
\left(\frac{\bar F(b)}{B}\right)_{E-L}=\frac{-1}{4\pi
b^{2}}\frac{\partial\bar E(b)}{\partial b}=\frac{\eta^{2}}{24\pi
b^{4}}(c_{1}\Lambda_{bet}+c_{2}\Lambda_{out})\left(1+\frac{1}{64\pi^5\eta^2}(\Lambda_{bet}^{-1}-
\Lambda_{out}^{-1})\zeta(2)\right),
\end{equation}
where $\zeta(2)$ is the Zeta function and Abel-Plana summation
formula has been used to regularize the infinite sum
$\sum_{i}\omega_i$. \\
Under conformal transformation, interaction
forces are given by
\begin{equation}\label{inteq}
\bar
P_{a}(a,b,a)=\frac{\eta^{2}\Lambda_{bet}}{3}P_{a}(a,b,a),\hspace{1cm}
\bar P_{b}(a,b,b)=\frac{\eta^{2}\Lambda_{bet}}{3}P_{b}(a,b,b).
\end{equation}
Therefore the total stress on the spheres due to boundary
conditions are obtained
$$
\frac{\bar F_{a}(a,b)}{A}=\frac{\bar F(a)}{A}-\bar P_{a}(a,b,a)=
$$
\begin{equation}\label{Ftot1}
\frac{\eta^{2}}{24\pi
a^{4}}(c_{1}\Lambda_{out}+c_{2}\Lambda_{bet})+\frac{1}{1536\pi^6a^4}(\Lambda_{out}^{-1}-
\Lambda_{bet}^{-1})(c_{1}\Lambda_{out}+c_{2}\Lambda_{bet})\zeta(2)-
\frac{\eta^{2}\Lambda_{bet}}{3}P_{a}(a,b,a),
\end{equation}
$$
\frac{\bar F_{b}(a,b)}{B}=\frac{\bar F(b)}{B}+\bar P_{b}(a,b,b)=
$$
\begin{equation}\label{Ftot2}
\frac{\eta^{2}}{24\pi
b^{4}}(c_{1}\Lambda_{bet}+c_{2}\Lambda_{out})+\frac{1}{1536\pi^6a^4}(\Lambda_{bet}^{-1}-
\Lambda_{out}^{-1})(c_{1}\Lambda_{bet}+c_{2}\Lambda_{out})\zeta(2)+
\frac{\eta^{2}\Lambda_{bet}}{3}P_{b}(a,b,b).
\end{equation}

Now, we obtain the pure effect of vacuum polarization due to the
gravitational field without any boundary conditions in Euclidean
region with the following metric
\begin{equation}
ds^{2}=-\Omega^2(\eta)[d\eta^{2}+\sum_{i=1}^{3}(dx^{i})^{2}].
\label{14}
\end{equation}
To this end, we calculate the renormalized stress tensor for the
massless scalar field in de Sitter space with Euclidean signature.
We use Eq.(\ref{15}), then we obtain
$\langle0|T^{\nu}_{\mu}[g_{kl}]|0\rangle|_{ren}=0$, and
$$
^{(1)}H^{\nu}_{\mu}=0,
$$
$$
^{(3)}H^{\nu}_{\mu}=\frac{3}{\alpha^4}\delta^{\nu}_{\mu}.
$$
Therefore
\begin{equation}
\langle0| T^{\nu}_{\mu}[\tilde g_{kl}]|0\rangle|_{ren}=\frac{1}{960
\pi^2 \alpha^4}\delta^{\nu}_{\mu}, \label{16}
\end{equation}
which is exactly the same result for the Lorentzian case
\cite{db}. The corresponding effective radial pressures for the
Euclidean (between) and Lorentzian (outside) regions with
$\alpha_{bet}$ and $\alpha_{out}$, due to pure effect of
gravitational vacuum polarization without any boundary condition,
are given respectively by
$$
P_{bet}^{E}=-\langle0|T^{r}_{r}[\tilde
g_{kl}]|0\rangle|_{ren}=-\frac{1}{960 \pi^2 \alpha^4_{bet}},
$$
$$
P_{out}^{L}=-\langle0|T^{r}_{r}[\tilde
g_{kl}]|0\rangle|_{ren}=-\frac{1}{960 \pi^2 \alpha^4_{out}}.
$$
The corresponding gravitational pressure on a single spherical shell
is then given by
\begin{equation}
P_g = P_{in}-P_{out} =
-\frac{1}{960\pi^{2}}\left(\frac{1}{\alpha_{in}^{4}}-\frac{1}{\alpha_{out}^{4}}\right).
\label{17}
\end{equation}
Therefore,gravitational pressures over each sphere, are given by
\begin{equation}\label{pgeq}
P_{g}(a)=\frac{-1}{8640\pi^{2}}(\Lambda_{out}^{2}-\Lambda_{bet}^{2}),\hspace{1cm}
P_{g}(b)=\frac{-1}{8640\pi^{2}}(\Lambda_{bet}^{2}-\Lambda_{out}^{2}).
\end{equation}
The total stress on each spherical shell due to pure gravitational
and boundary effects , is then obtained
$$
P_{tot}(a)=\frac{\eta^{2}}{24\pi
a^{4}}(c_{1}\Lambda_{out}+c_{2}\Lambda_{bet})+\frac{1}{1536\pi^6a^4}(\Lambda_{out}^{-1}-
\Lambda_{bet}^{-1})(c_{1}\Lambda_{out}+c_{2}\Lambda_{bet})\zeta(2)
$$
\begin{equation}\label{ptot1}
-
\frac{1}{8640\pi^{2}}(\Lambda_{out}^{2}-\Lambda_{bet}^{2})-\frac{\eta^{2}\Lambda_{bet}}{3}P_{a}(a,b,a),
\end{equation}
$$
P_{tot}(b)=\frac{\eta^{2}}{24\pi
b^{4}}(c_{1}\Lambda_{bet}+c_{2}\Lambda_{out})+\frac{1}{1536\pi^6b^4}(\Lambda_{bet}^{-1}-
\Lambda_{out}^{-1})(c_{1}\Lambda_{bet}+c_{2}\Lambda_{out})\zeta(2)
$$
\begin{equation}\label{ptot2}
-\frac{1}{8640\pi^{2}}(\Lambda_{bet}^{2}-\Lambda_{out}^{2})+\frac{\eta^{2}\Lambda_{bet}}{3}P_{b}(a,b,b).
\end{equation}

\section{Discussion}

We may discuss on different possible cases for the total stresses.
Let us first consider $P_{tot}(a)$ and assume $\Lambda_{bet}
> \Lambda_{out}$ with
\begin{equation}\label{cond1}
c_{1}\Lambda_{out}+c_{2}\Lambda_{bet}>0,
\end{equation}
which leads the three first terms in $P_{tot}(a)$ to be positive.
Also, noting that $P_{a}(a,b,a)<0$, the forth term in Eq.
(\ref{ptot1}) is always positive, therefore the total pressure
$P_{tot}(a)$ on the inner shell is repulsive.

For the case $\Lambda_{out}>\Lambda_{bet}$ the first and forth
time dependent terms are positive while the second and third
constant terms are negative. Therefore the total pressure
$P_{tot}(a)$ may be either negative or positive. Given
$P^{tot}(a)<0$ initially ($\eta$=0), the first and forth terms
will dominate after some time and the pressure changes to be
positive, namely a repulsive one. The situation for $P_{tot}(b)$
is as follows. Consider
\begin{equation}\label{cond1}
c_{1}\Lambda_{bet}+c_{2}\Lambda_{out}>0.
\end{equation}
For $\Lambda_{bet} > \Lambda_{out}$, all terms but the first one
are negative. At $\eta=0$, the pressure $P_{tot}(b)$ is negative,
but at sufficiently late times there is a competition between the
first positive and the forth negative time dependent terms,
according to which the outer sphere will experience a repulsive or
an attracting force, respectively. For the case
$\Lambda_{out}>\Lambda_{bet}$, all the terms but the forth one are
positive. At $\eta=0$, the pressure $P_{tot}(b)$ is positive, but
at sufficiently late times there is again a competition between
the first positive and the forth negative time dependent terms,
according to which the outer sphere will experience a repulsive or
an attracting force, respectively.

Now, we consider the case
\begin{equation}\label{cond3}
c_{1}\Lambda_{out}+c_{2}\Lambda_{bet} <0,
\end{equation}
with $\Lambda_{bet}> \Lambda_{out}$. In this case, the first and
second terms are negative whereas the third and forth terms in
$P_{tot}(a)$ are positive. At $\eta=0$, the pressure $P_{tot}(a)$
may be positive or negative and at late times there is a
competition between the first negative and the forth positive time
dependent terms, according to which the outer sphere will
experience an attracting or a repulsive force. For $\Lambda_{out}>
\Lambda_{bet}$, the first and third terms are negative and the
second and forth terms are positive. At $\eta=0$, the pressure
$P_{tot}(a)$ may be negative or positive and at late times there
is a competition between the first negative and the forth positive
time dependent terms, according to which the outer sphere will
experience an attracting or a repulsive force.

In this case, the pressure $P_{tot}(b)$ behaves as follows.
Consider
\begin{equation}\label{cond1}
c_{1}\Lambda_{bet}+c_{2}\Lambda_{out}<0.
\end{equation}
For $\Lambda_{bet} > \Lambda_{out}$, all the terms but the second
one are negative. At $\eta=0$, the pressure $P_{tot}(b)$ may be
positive or negative according to the second and third terms. For
late times, however, the pressure becomes negative leading to an
attracting force. For $\Lambda_{bet} < \Lambda_{out}$, all the
terms but the third one are negative. At $\eta=0$, the pressure
$P_{tot}(b)$ may be negative or positive according to the second
and third terms. For late times, however, the pressure becomes
negative leading to an attracting force.

    \section{Conclusion}
    We have revisited two concentric spherical shells in a signature changing de Sitter
    metric conformally coupled to a massless scalar field, satisfying Dirichlet boundary
    conditions, and different cosmological terms inside and outside the spheres.
    We have obtained time independent extra terms for
    the total pressures on each sphere. These terms vanish for the
    special case of identical cosmological constants, as it happens for the gravitational
    pressures. For different cosmological constants, however, the
    introduction of an Euclidean signature between the spheres
    leads to an extra constant term on each sphere.

    Given a false vacuum $\Lambda_{bet}$ between and a true
    vacuum $\Lambda_{out}$ outside the spheres, namely $\Lambda_{bet}>
    \Lambda_{out}$, the constant terms arising from signature
    change contribute positively ( negatively ) to the total pressure for the inner ( outer ) sphere,
    compared with the pure Lorentzian case. Therefore, in this
    case there is an extra attraction between the two spheres. This
    attraction leads the Euclidean region to become smaller and be
    replaced by Lorentzian region.

    On the contrary, given a true vacuum $\Lambda_{bet}$ between and a
    false vacuum $\Lambda_{out}$ outside the spheres, namely $\Lambda_{out}>
    \Lambda_{bet}$, the constant terms arising from signature
    change contribute negatively ( positively ) to the total pressure for the inner ( outer ) sphere,
    compared with the pure Lorentzian case. Therefore, in this
    case there is an extra repulsion between the two spheres. This
    repulsion leads the Lorentzian regions to become smaller and be
    replaced by Euclidean region.

    We then come to the conclusion that in both cases the region
    endowed by the true vacuum is capable of extending to
    those regions of false vacuum.
    This may be of particular importance in the study of inflationary models of the early
    universe. If we assume there were some regions of Euclidean
    signature within spherical layers ( endowed by a false
    vacuum ) inside the bubbles with Lorentzian signature ( endowed by a true vacuum ),
    then these regions tend to be removed and replaced by Lorentzian regions, due to the Casimir effect.

  \vspace{3mm}

\section*{Acknowledgment}

This work has been supported by Research Institute for Astronomy and
Astrophysics of Maragha.

\end{document}